\documentclass[useAMS,usenatbib]{mn2e}
\usepackage{graphicx,verbatim,color}
\usepackage[authoryear]{natbib}

\renewcommand{\textcolor}[2]{#2}

\title[Activity-induced RV variations from photometry]{A simple method to estimate radial velocity variations due to stellar activity using photometry}

\author[S. Aigrain et al.]{S. Aigrain$^{1}$\thanks{E-mail: suzanne.aigrain@astro.ox.ac.uk}, F. Pont$^{2}$, S. Zucker$^{3}$\\ 
  $^{1}$Sub-department of Astrophysics, Department of Physics, University of Oxford, Oxford OX1 3RH, UK\\ 
  $^{2}$Astrophysics Group, School of Physics, University of Exeter, Exeter EX4 4QL, UK\\ 
  $^{3}$Department of Geophysics and Planetary Sciences, Raymond and Beverly Sackler Faculty of Exact Sciences,\\
  $^{~}$Tel Aviv University, Tel Aviv 69978, Israel}

\begin{document}

\date{Accepted \ldots; Received \ldots; in original form \ldots}

\pagerange{\pageref{firstpage}--\pageref{lastpage}} \pubyear{\ldots}

\maketitle

\label{firstpage}

\begin{abstract}
We present a new, simple method to predict activity-induced radial velocity variations using high-precision time-series photometry. It is based on insights from a simple spot model, has only two free parameters (one of which can be estimated from the light curve) and does not require knowledge of the stellar rotation period. We test the method on simulated data and illustrate its performance by applying it to MOST/SOPHIE observations of the planet host-star HD\,189733, where it gives almost identical results to much more sophisticated, but highly degenerate models, and synthetic data for the Sun, where we demonstrate that it can reproduce variations well below the m\,s$^{-1}$ level. We also apply it to Quarter 1 data for Kepler transit candidate host stars, where it can be used to estimate RV variations down to the 2--3\,m\,s$^{-1}$ level, and show that RV amplitudes above that level may be expected for approximately two thirds of the candidates we examined.
\end{abstract}


\section{Introduction}

Stellar activity induces brightness and line-shape variations which can mimic planetary signals and hinder the detection and/or characterisation of the latter. In particular, in radial velocity (RV) surveys, many stars display intrinsic variability which is attributed to activity, and which occurs on timescales similar to the planetary signals of interest. There is thus significant interest in characterising the level of activity-induced RV variability in stars targeted by planet surveys, and in developing tools to distinguish between the latter and planetary signals. 

The simplest, widely used method to deal with activity-induced variability in RV surveys is to add a `jitter' term in quadrature to the RV uncertainties before searching for planetary signals. \citet{wri05} proposed an empirical relation to predict the magnitude of this jitter term from a star's activity level, $B-V$ colour and absolute magnitude, calibrated on 450 targets from the California and Carnegie planet search, but this relation is far from tight. More recently, \citet{isa+11} determined chromospheric activity levels for over 2500 target stars from the same survey, and compared them to the RMS scatter of their RV variations. Again, there is clearly a relationship between the two -- in particular, the lower envelope of the RV scatter increases for increasing levels of activity -- but there is also a wide range of RV scatters for a given spectral type and activity level. Furthermore, the `jitter' formalism is limited, because it treats the activity signal as an independent, identically distributed Gaussian noise process. Activity-induced RV signals arise from the rotational modulation and intrinsic evolution of magnetized regions, and are thus naturally correlated in time, often quasi periodic, and non-stationary. Therefore the impact of variability-induced RV signals on planet detection will generally be much more severe than that of a random “jitter” of the same mean amplitude.

For individual stars, somewhat more sophisticated approaches are in common use, mostly making use of chromospheric activity indicators, such as excess flux in the cores of the H$\alpha$ and Ca {\sc ii} H \& K lines, of measurements of the degree of asymmetry of the spectral lines, such as the bisector span of the cross-correlation function (CCF) between the stellar spectrum and a template, which is often used to derive the RV itself. The most obvious step is to check for periodic modulation in these indicators, which can reveal that a suspected planetary signal is in fact due to activity \citep[see e.g.][]{que+01,bon+07}. The correlation between RVs and bisector span can also be used to \emph{correct} for the effect of activity at the few m\,s$^{-1}$ level \citep{mel+07,boi+09}.  However, this correlation is highly dependent on the spot distribution and stellar inclination \citep{boi+11} and is not always present. In \citet{pon+11}, we developed another method, which consists in modeling the variations of the stellar brightness and the CCF bisector span using a spot model, to predict the activity-induced variations. Spot models are very degenerate: many spot distributions with widely different numbers of spots and spot parameters fit the data equally well, so the predictions of many models must be averaged. One way to address this degeneracy is to use maximum entropy regularisation, as was done by \citet{lan+07}, to model the Sun's total irradiance variations. Recently, \citet{lan+11} applied this method to the planet-host star HD\,189733, using photometric observations performed by the MOST satellite to predict the activity signal in simultaneous SOPHIE RV observations, and obtained slightly better results than \citet{boi+09}, who had previously analysed the same dataset using the bisector span de-correlation method.

The data from space-based transit surveys such as CoRoT and Kepler contain a wealth of information about stellar activity, but the spot-modeling approaches above are computationally intensive to be applied systematically to many light curves. They are also dependent on fairly detailed a-priori knowledge of the star being modeled, and in particular of its rotation period. In this paper, we present a new, much simpler method to predict the activity-induced RV variations for a given star from its photometric variations, which can be applied to stars whose rotation period is not known. This is intended primarily for statistical purposes: to characterise the overall RV variability properties of a sample of stars, and to select the best targets for RV follow-up, for example. However, as Kepler continues to observe the candidate  planets it has identified while they are being followed up from the ground, our method may also prove useful in reducing contamination of the RV data by activity signals in individual cases.

Our method is based on a very simple spot model, which we introduce in Section~\ref{sec:spot_model}. We then make use of a relationship between the photometric and RV signatures of individual spots to formulate a means of simulating the RV variations based on the light curve only, as outlined in Section~\ref{sec:method}. In Section~\ref{sec:tests}, we test the method on simulated data and give three example applications: the SOPHIE/MOST observations of HD\,189733b already used as a test case by \citet{boi+09} and \citet{lan+11}, total irradiance and RV variations of the Sun synthesized by \citet{meu+10b}, and Kepler quarter 1 light curves for a subset of the planetary candidates published by \citet{bor+11}. We present our conclusions and future plans for applying the method to other datasets, in Section~\ref{sec:conc}.

\section{A simple spot model}
\label{sec:spot_model}

\citet{dor87} provided analytic expressions to model the photometric
signature of a circular spot on a rotating stellar surface, accounting
for limb-darkening using a single-coefficient, linear limb-darkening
law. It is possible to expand these to also model the RV signature of
such a spot, and to account for both dark spots and bright faculae by
allowing the contrasts and limb-darkening coefficients to change
sign. However, the resulting algebra is relatively cumbersome. Here we
seek a simpler model, whose mathematical expressions are simple enough
to afford a more direct insight into the dependency of the photometric
and RV signatures of active regions on the various parameters, while
retaining as much realism as possible. In order to achieve this, we
make a number of simplifications, the impact of which we will test by
comparing our results to data simulated using the more complete
\citet{dor87} formalism.

\subsection{Photometric signature of a point-like spot}

To model the photometric signature of dark spots, we
  assume that the spots are small, i.e. that $\alpha \ll 1$, where
  $\alpha$ is the spot's angular radius on the surface of the
  star. This allows us to ignore projection effects within a spot, and
  obviates the need to assume a particular shape for the spots. It
  also enables us, when considering multiple spots, to assume that
  they never overlap. Additionally, we ignore limb-darkening. This
  alters the photometric signature of dark spots only slightly, since
  both the spot and the un-spotted photosphere are darker towards the
  limb than they are near the centre of the stellar disk.

  Under these conditions, the relative drop in flux due to a single
  point-like, dark spot rotating on the stellar surface is
\begin{equation}
  \label{eq:F}
  F(t) = f ~ {\rm MAX} \left\{  \cos \beta(t); 0 \right\},
\end{equation}
where $f = 2 \textcolor{red}{(1-c)} (1 - \cos \alpha)$ represents the
relative flux drop for a spot at the disk centre, $c$
is the contrast ratio between the spot and the un-spotted photosphere,
and $\beta(t)$ is the angle between the spot normal
and the line of sight. 
This angle is given by
\begin{equation}
  \label{eq:beta}
  \cos \beta(t) = \cos \phi(t) \, \cos \delta \, \sin i + \sin \delta \cos i,
\end{equation}
where $i$ is the stellar inclination (the angle between the star's
rotation axis and the line of sight), $\delta$ is the
latitude of the spot relative to the star's rotational equator, and
$\phi(t)$ is the phase of the spot relative to the line of sight (see
Appendix~\ref{sec:append} for details). The latter is of course
$\phi(t) \equiv 2 \pi t / P_{\rm rot} + \phi_0$, where $P_{\rm rot}$
is the star's rotation period and $\phi_0$ is the longitude of the
spot (i.e. we take the stellar meridian to be aligned with the line of
sight at $t=0$). The observed stellar flux is then simply
\begin{equation}
  \Psi(t) = \Psi_0 \left[ 1 - F(t) \right],
\end{equation}
where $\Psi_0$ is the flux in the absence of spots.

By allowing $c$ to \textcolor{red}{exceed one}, one could use the
equations above to model the signature of bright spots such as
faculae. However, faculae on the Sun are limb-brightened \citep[see
e.g.][and references therein]{lan+07,meu+10a}, and any model that
ignores the limb-angle dependence of the contrast will not reproduce
their photometric signature well. Thus we have opted not to include
faculae in our model. Fortunately, the latter typically have low
contrast, and hence the missing photometric effect tends to be
small. \textcolor{red}{Comparative studies of Sun-like stars by
  \citet{rad+98} and \citet{loc+07} also suggest that faculae are less
  important in stars more active than the Sun.}

\subsection{RV signature}
\label{sec:defRV}

\textcolor{red}{The most important effect of the spot in RV is due to
  the fact that it suppresses the flux emitted by a portion of the
  rotating stellar disk, thus introducing a perturbation to the
  disk-averaged RV. Provided $c$ is not close to one, this perturbation
  can be estimated simply by multiplying the projected area of the
  spot, which is given by $F(t)$, by the RV of the stellar surface at
  the location of the spot:}
\begin{equation}
  \label{eq:RV}
  \Delta RV_{\rm rot}(t) = \textcolor{red}{-} F(t) \, V_{\rm eq} \, \cos \delta \sin \phi(t) \, \sin i,
\end{equation}
where $V_{\rm eq} = 2 \pi R_\star / P_{\rm rot}$ is the equatorial
rotational velocity of the star, and $R_\star$ the stellar
radius. Differential rotation can be included in this formalism by
allowing $V_{\rm eq}$ (or equivalently $P_{\rm rot}$) to vary as a
function of the spot \textcolor{red}{latitude}.

Spots tend to be associated with magnetized areas which, while they
have very limited photometric contrast, are much more extended
spatially. These do have an important impact in RV, because convection
is partially suppressed within them, leading to a reduction in the
convective blue-shift \citep[see][and references
therein]{meu+10a,meu+10b}. Within our simplified formalism, it is
possible to approximate the resulting RV perturbation as
\begin{equation}
  \Delta RV_{\rm c}(t) = \textcolor{red}{+} F(t) \, \delta V_{\rm c} \, \kappa \, \cos \beta(t)
\end{equation}
where $\delta V_{\rm c}$ is the difference between the convective
blue-shift in the unspotted photosphere and that within the magnetized
area, and $\kappa$ is the ratio of this area to the spot surface
(typically $\gg 1$). The total RV signature of the spot and associated
magnetised area is then simply
\begin{equation}
  \Delta RV(t) = \Delta RV_{\rm rot}(t) + \Delta RV_{\rm c}(t).
\end{equation}

\subsection{Examples}

\begin{figure}
  \centering
  \includegraphics[width=\linewidth]{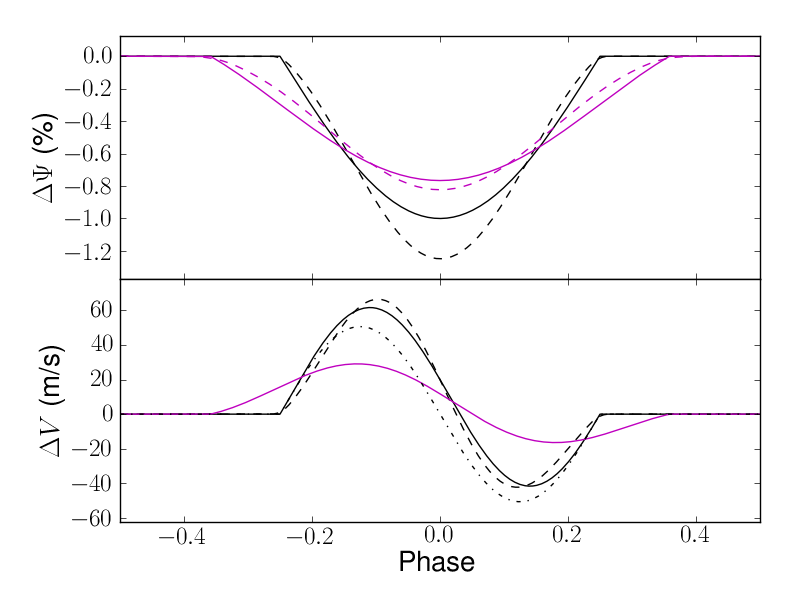}
  \caption{\textcolor{red}{Simulated photometric and RV signatures of
      a single spot (top and bottom respectively). The solid black
      line shows the output of our simple model for a fairly large,
      dark, equatorial spot ($c=0$, $\alpha = 10\,^\circ$,
      $\delta=0$) on a star with $P_{\rm rot}=5$\,days,
      $i=90\,^\circ$, $\delta V_{\rm c}=200$\,m\,s$^{-1}$ and
      $\kappa=10$ (see text for details). The solid cyan line is the
      same, but for a higher latitude spot on an inclined star
      ($\delta=60\,^\circ$, $i=70\,^\circ$). For comparison, the
      dashed lines show the same spots modeled with the formalism of
      \protect\citet{dor87} (stellar linear limb-darkening parameter
      $u_\star=0.5$). For simplicity, we have omitted the Dorren
      formalism for the high-latitude case in the bottom
      panel. Instead, the black dash-dot line shows the equatorial
      spot simulated with our simple model, but without the convective
      blue-shift effect ($\delta V_{\rm c}=0$). }}
  \label{fig:singlespot}
\end{figure}

\begin{figure}
  \centering
  \includegraphics[width=\linewidth]{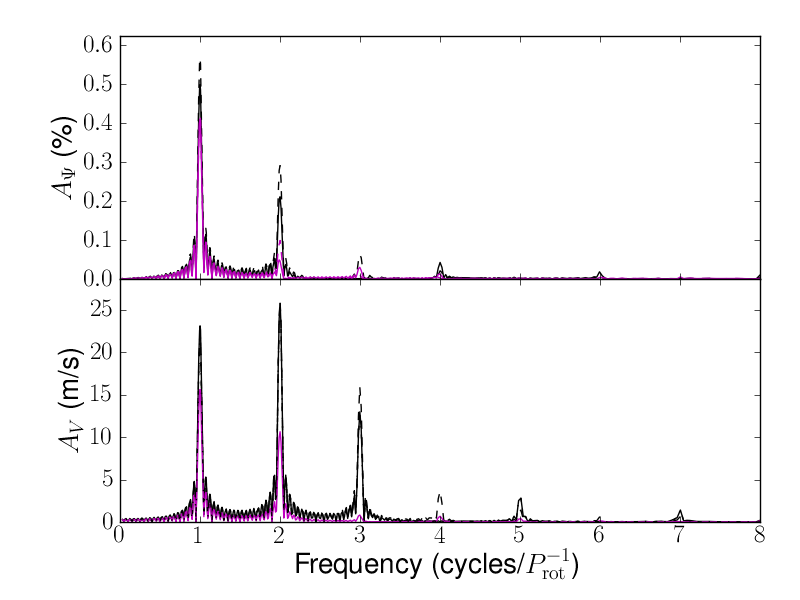}
  \caption{Amplitude spectra for the light and RV curves shown in
    \protect\ref{fig:singlespot}, using the same colour-coding. These
    spectra were computed from time-series lasting $5 \, P_{\rm
      rot}$. Frequencies are expressed in units of inverse rotation
    periods.}
  \label{fig:singlespot_per}
\end{figure}

Figure~\ref{fig:singlespot} shows light and RV curves simulated using
this simple model for an equatorial spot (solid black line) and a
high-latitude spot on an inclined star (solid cyan line). The dotted
grey line in the bottom panel shows the equatorial spot case without
the convective blue-shift suppression term. Also shown for comparison
is the same equatorial spot modeled with the more sophisticated
formalism of \citet{dor87}, who gives analytical expressions
for a circular spot of finite size on a limb-darkened photosphere.
Figure~\ref{fig:singlespot_per} shows the corresponding amplitude
spectra\footnote{Throughout this paper, we use amplitude spectra
  computed by linear least-squares fitting of a sinusoid with free
  zero-point, amplitude and phase at each frequency. This is akin to
  the generalised periodogram of \protect\citet{zec+09} but expressed
  in units of amplitude rather than $\chi^2$ reduction.}, which we use
to evaluate the impact of the simplifications we have made on the
frequency content of the simulated light and RV curves.

In all cases, the amplitude spectrum of the light curve is dominated
by the rotational frequency, as one might expect. There is also signal
at the \textcolor{red}{second, fourth, sixth and higher even-numbered
  harmonics} $\nu = 2n/P_{\rm rot}$, where $n=1,2,3$,\ldots, although
the amplitude decreases rapidly with $n$. On the other hand, there is
essentially no signal at \textcolor{red}{odd-numbered harmonics}
$\nu=(2m+1)/P_{\rm rot}$, where $m=1,2,3$,\ldots. As previously noted
by \citet{boi+09,boi+11}, the RV signature is dominated by the
\textcolor{red}{first three harmonics of the rotational frequency},
with additional signal at \textcolor{red}{odd-numbered harmonics},
although at much lower amplitude. The differences between the
distribution of power at higher harmonics in photometry and in RV
arise because the photometric signal is maximised when the spot is
face on, which occurs once per disk-crossing, but the RV signal is
maximised when the spot longitude is $45^\circ$, which occurs twice
per disk crossing.

Changing the spot latitude and stellar inclination can substantially
alter the light and RV curves, as it changes the fraction of the
rotation cycle over which the spot remains in- or out-of-view, and the
rotational velocity of the occulted parts of the stellar disk. This
results in significant changes in the relative amounts of power at the
\textcolor{red}{different harmonics of the rotational frequency}, in
some cases entirely suppressing \textcolor{red}{all but the
  fundamental}, or on the contrary giving rise to relatively large
amplitudes at high-$n$ \textcolor{red}{harmonics}.  Projection effects
over the area of the spot, and limb-darkening (both of which are
accounted for in the \citealt{dor87} formalism, but not in our simple
model), alter the shape and maximum amplitude of the light curve, and
hence the balance of power between the fundamental and
\textcolor{red}{second harmonic}, but not in a very substantial way
(except for extremely large spots). The effect on the RV signature is
even smaller.

The convective blue-shift causes the RV perturbation to be biased
upwards and to depart from an exact sinusoidal shape.  As $\Delta
RV_{\rm c}(t)$ is proportional to $F^2$, the power of the convective
component of the RV signature is concentrated at the rotational
frequency and its first harmonic. Except for extreme cases, the effect
of convective blue-shift on the frequency content of the RV curve
remains minor.

\section{The $FF'$ method}
\label{sec:method}

\subsection{Relationship between photometric and RV signatures}

Considering the equations of our spot model, as presented in
Section~\ref{sec:spot_model}, it is interesting to note that
\begin{eqnarray}
  \dot{F}(t) & = & - \textcolor{red}{f} \, \sin \phi(t) \, \dot{\phi}(t) \, \cos \delta \, \sin i \nonumber\\ 
  & = & - \textcolor{red}{f} \, \sin \phi(t) \, \cos \delta \, \sin i \,
  \frac{2 \pi}{P_{\rm rot}}. 
\end{eqnarray}
Therefore the expression for the RV signature of spots can be re-written:
\begin{equation}
\Delta RV_{\rm rot}(t) = - F(t) \, \dot{F}(t) \, R_\star / \textcolor{red}{f}.
\end{equation}
This can be estimated directly from the light curve, as
\begin{equation}
F(t) = 1 - \frac{\Psi(t)}{\Psi_0} \hspace{1cm} {\rm and} \hspace{1cm}
\dot{F}(t) = - \frac{\dot{\Psi}(t)}{\Psi_0},
\end{equation}
hence
\begin{equation}
\label{eq:ff'}
\Delta RV_{\rm rot}(t) = \frac{\dot{\Psi}(t)}{\Psi_0} \, \left[ 1 - \frac{\Psi(t) }{\Psi_0} \right]  \, \frac{R_\star}{\textcolor{red}{f}}.
\end{equation}
Similarly, the convective blue-shift effect is given by
\begin{equation}
\Delta RV_{\rm c}(t) = + F^2(t) \, \delta V_{\rm c} \, \kappa / \textcolor{red}{f},
\end{equation}
which can be estimated from the light curve as 
\begin{equation}
\Delta RV_{\rm c}(t) = + \left[ 1 - \frac{\Psi(t)}{\Psi_0} \right]^2 \, \frac{\delta V_{\rm c} \, \kappa}{\textcolor{red}{f}}.
\end{equation}

The above expressions provide a means to simulate activity-induced RV
variations based on a well-sampled light curve, \emph{without knowing
  the rotation period}. As this method uses the light curve and its
own derivative, we refer to it as the $FF'$ method.

The effect of multiple spots simultaneously present on the stellar
surface is additive: the observed flux modulation is the sum of the
contributions from individual spots, and similarly for RV. However,
strictly speaking, the reasoning behind the $FF'$ method cannot be
extended to multiple spots, since the direct relationship between the
RV signature of each spot and the light curve breaks
down. \textcolor{red}{Therefore, we expect the $FF'$ method to perform
  best when a single active region dominates the visible
  hemisphere. When multiple active regions are present, we are
  essentially making a first order approximation: therefore the $FF'$
  should still reproduce the dominant features of the RV variations,
  but will necessarily be less accurate, particularly on short
  timescales. W}e shall test the extent to which this is the case
using both observed and simulated data.

\subsection{Practical implementation}
\label{sec:practical}

To compute the RV signal from the light curve, one must estimate
$R_\star$, $\Psi_0$, $f$ and $\delta V_{\rm c} \,
\kappa$. We will assume that $R_{\star}$ is known at least
approximately, which is generally the case. If the
  distribution of active regions is relatively smooth, so that there
  are always several active regions on both of the visible and the
  hidden hemisphere, then the maximum of the observed light curve,
  $\Phi_{\rm max}$, will be offset from $\Psi_0$. In that case, one
  may expect a direct scaling between this offset and the amount of
  variability in the light curve scatter. We adopt the expression
\begin{equation}
\label{eq:psi0}
  \Psi_0 \approx \Phi_{\rm max} + k \sigma.
\end{equation}
where $\sigma$ is the light curve scatter and $k$ is a free
parameter. We use the scatter rather than, say, the peak-to-peak
amplitude, because -- except for nearly pole-on stars, whose
variability will in any case be minimal -- active regions which are
always visible are necessarily relatively near the limb, whereas the
lowest excursions in the light curves are presumably caused by active
regions which come close to the centre of the visible disk. Initial
tests regarding the $k$ on a number of simulated test cases and on the
observations of HD\,189733, which are described in
Section~\ref{sec:hd189}, suggested that $k=1$ is a suitable value for
relatively active stars, and this is the value we use by
default. However, when enough data is available to constrain a free
parameter, it is advisable to fit for either $k$ or $\Psi_0$ itself.
Once $\Psi_0$ is determined, the largest observed departure from it
provides a fairly good estimate of the spot coverage:
\begin{equation}
  f\approx \frac{\Psi_0 - \Phi_{\rm min}}{\Psi_0}, 
\end{equation}
where $\Phi_{\rm min}$ is the minimum observed flux. Again, this
scaling relation performed well in initial tests on simulated data and
observations of HD\,189733b. Even in data-rich situations, it is
rarely helpful to fit for both $f$ and $\Psi_0$ as the two parameters
are mutually degenerate.

For well-sampled light curves, the derivative of the flux can be
estimated directly from the difference between consecutive data
points. This procedure is highly sensitive to high-frequency noise, so
the light curve must be smoothed first. One possibility is to do this
using the non-linear filter of \citet{aig+04} with a smoothing length
approximating a tenth of the rotation period (or, when the latter is
not known, of the dominant periodicity in the light curve). For an
example application using this filter, see
Section~\ref{sec:hd189}. When the time-sampling is not sufficient, an
alternative is to model the light curve using Gaussian process
regression (see Section~\ref{sec:sun} for an application, and
Appendix~\ref{app:GP} for more details on Gaussian process
regression).

\section{Tests and applications}
\label{sec:tests}

\subsection{Photometry is insensitive to certain spot distributions}

\begin{figure}
  \centering
  \includegraphics[width=\linewidth]{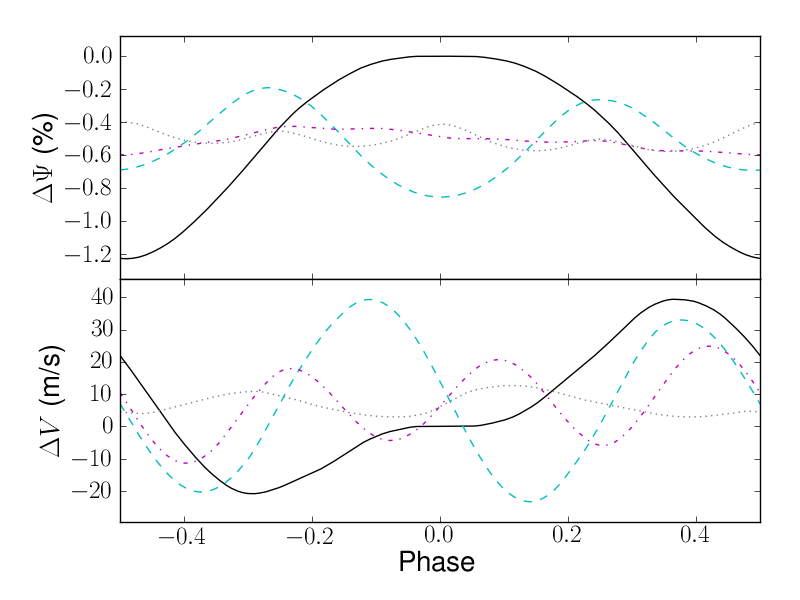}
  \caption{Photometric and RV signatures of spot distributions
    matching low-order multipoles with 1, 2, 3 and 4 nodes along
    the equator (solid black, dashed cyan, dash-dot magenta and dotted
    grey line respectively). The spot distributions are uniform in
    $\sin(\delta)$. The stellar parameters are identical to the
    fiducial values used in Figure~\protect\ref{fig:singlespot}.}
  \label{fig:multipoles}
\end{figure}
 
The $FF'$ method rests on the assumption that the photometric signal
contains sufficient information to adequately predict the RV
signal. We test this assumption here by examining the relative
amplitudes and shapes of photometric and RV signals from spot
distributions approximating low-order multipoles in the longitudinal
direction.  We simulated light and RV curves for stars with 200 spots,
each with $c=0$ and $\alpha=0.01^\circ$. The spot longitudes
$\phi_0$ were drawn at random from the desired distribution, and their
latitudes were taken from a uniform distribution in
$\sin(\delta)$. All the parameters of each spot were fixed (no spot
evolution) and all spots shared the same rotation rate (no
differential rotation).

The results of these tests are shown in
Figure~\ref{fig:multipoles}. While both photometric and RV signals are
sensitive to dipole (black line) and quadrupole (dashed cyan line)
configurations, the photometric signal is only very marginally
sensitive to higher order odd-numbered multipoles (e.g.\ order $3$,
dash-dot magenta line), which do give rise to some RV variation. This
can be understood intuitively as a consequence of the symmetry of the
problem.  The same phenomenon has already been noted by \citet{cow+08}
in the context of phase-function mapping of exoplanets: sinusoidal
maps with odd order have no photometric phase function
signature. Conversely, the RVs are not very sensitive to higher order
even-numbered multipoles (e.g.\ order $4$, dotted grey line)
configurations, although this is less noticeable. This implies that
any attempt to simulate activity-induced RV variations based on
photometry -- whatever the method -- is likely to over-estimate the
signal at high-$n$ (\textcolor{red}{even-numbered) harmonics}, and underestimate that at
high-$m$ (\textcolor{red}{odd-numbered) harmonics}.

\subsection{Simulations using multiple, evolving active regions}
\label{sec:sim}

\begin{figure*}
  \centering
  \includegraphics[width=0.3\linewidth]{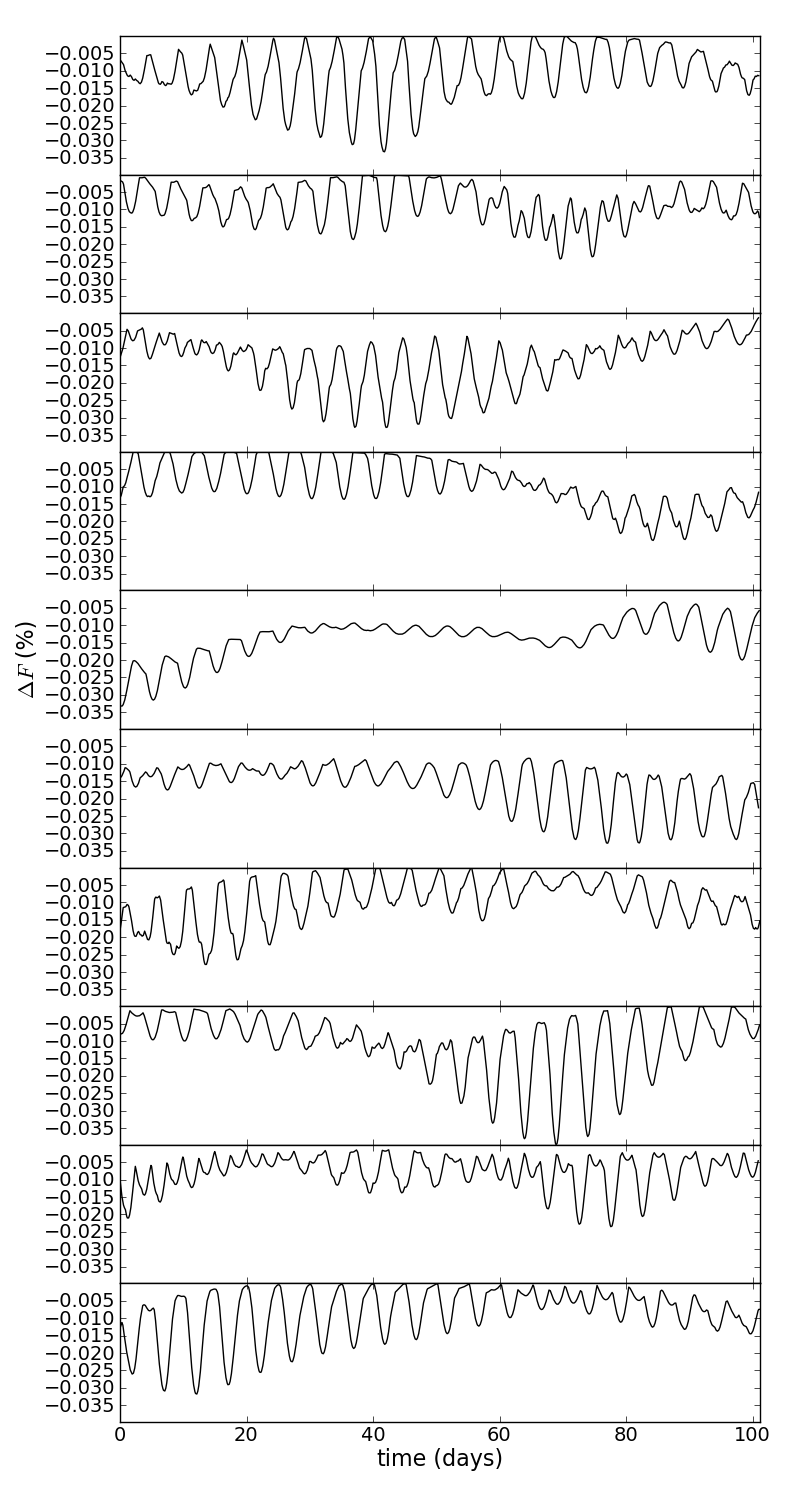}
  \includegraphics[width=0.3\linewidth]{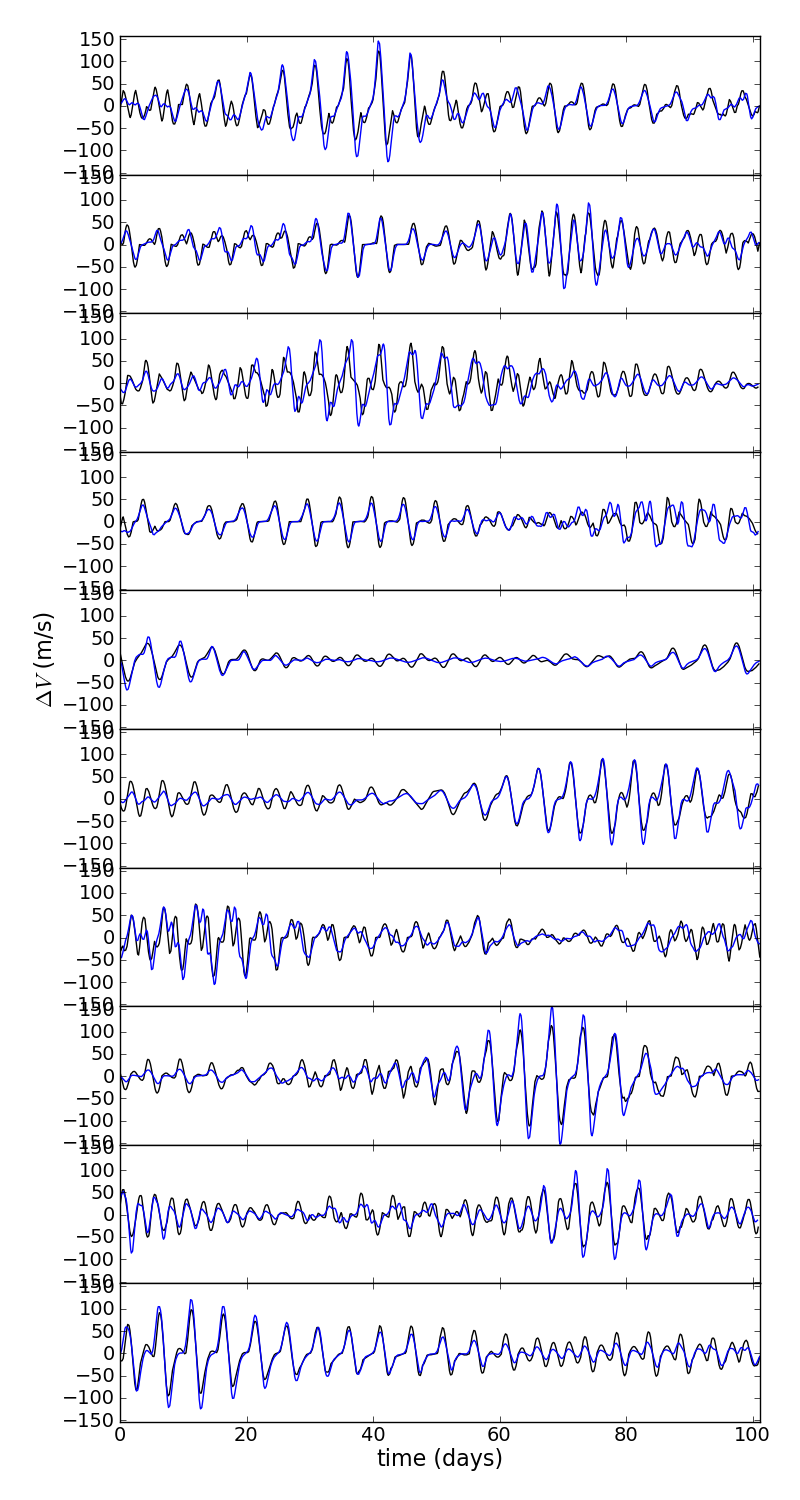}
  \includegraphics[width=0.3\linewidth]{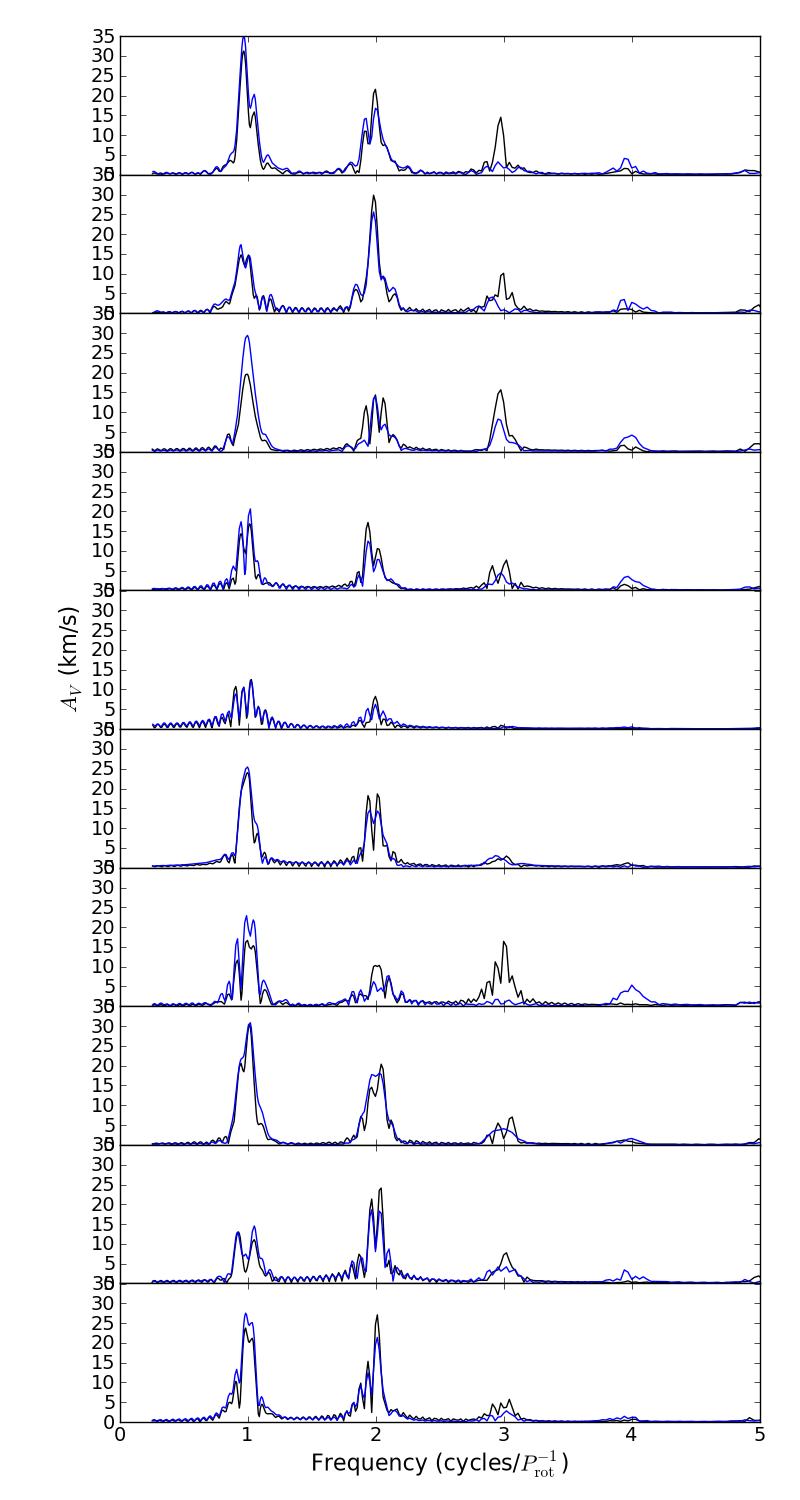}
  \caption{Simulation examples: flux variations simulated with the spot model (left), RV variations (middle) produced by the spot model (blue) and by the $FF'$ method applied to the flux data shown in the left panel (black), and corresponding periodograms (right). These 10 examples were selected at random from the 100 simulations with 20 spots.}
  \label{fig:simres_comp}
\end{figure*}

\begin{figure*}
  \centering
  \includegraphics[width=0.3\linewidth]{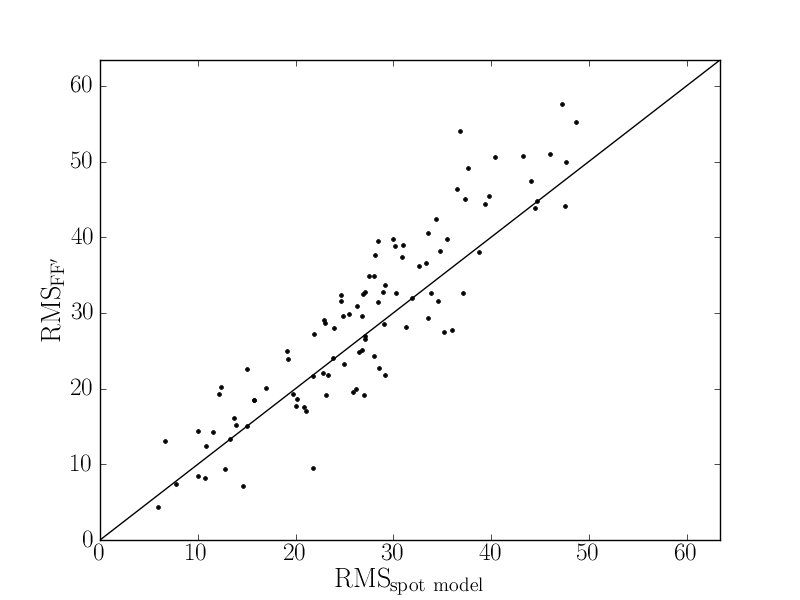}
  \includegraphics[width=0.3\linewidth]{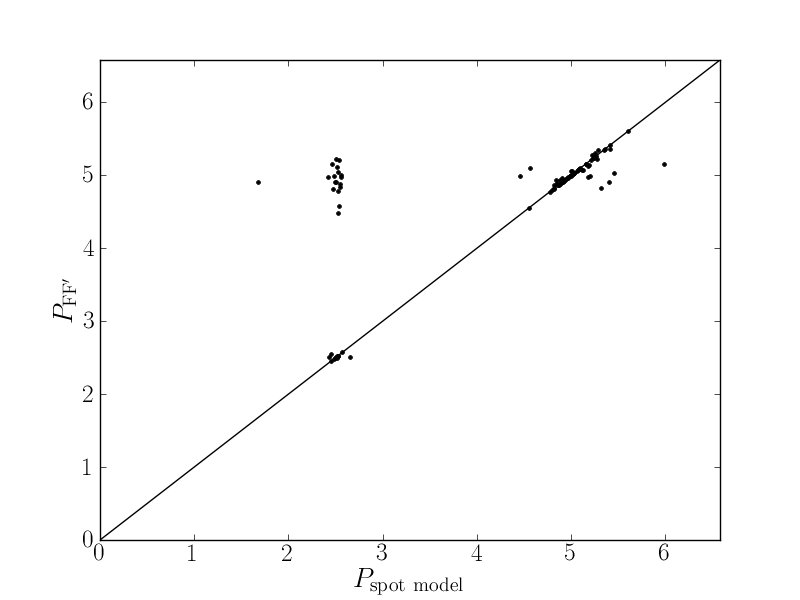}
  \includegraphics[width=0.3\linewidth]{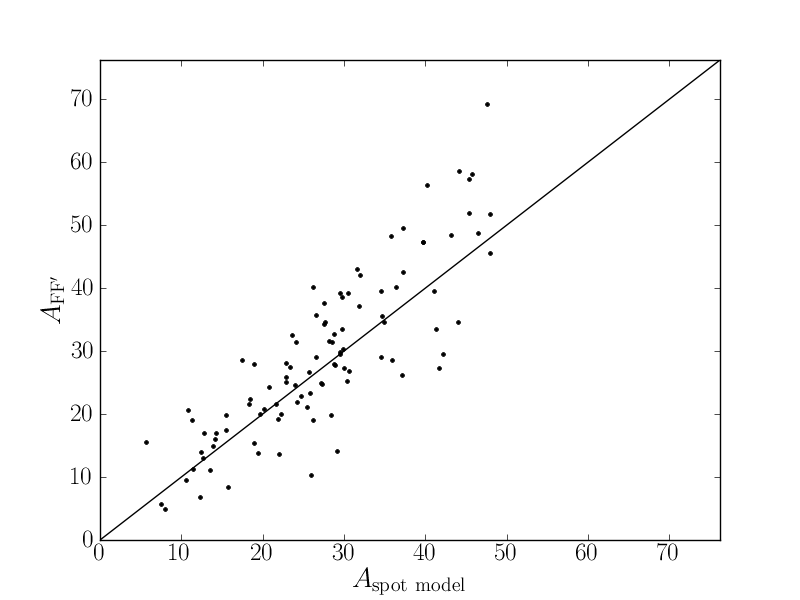}
  \includegraphics[width=0.3\linewidth]{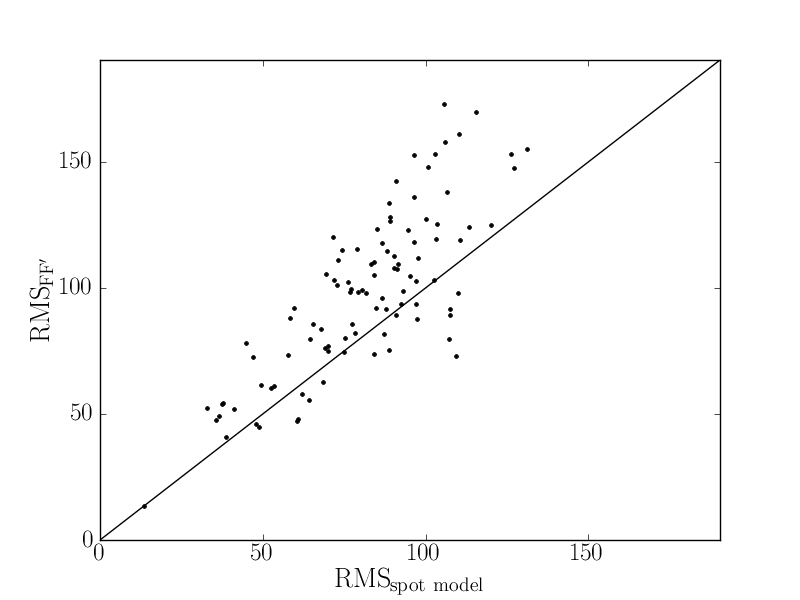}
  \includegraphics[width=0.3\linewidth]{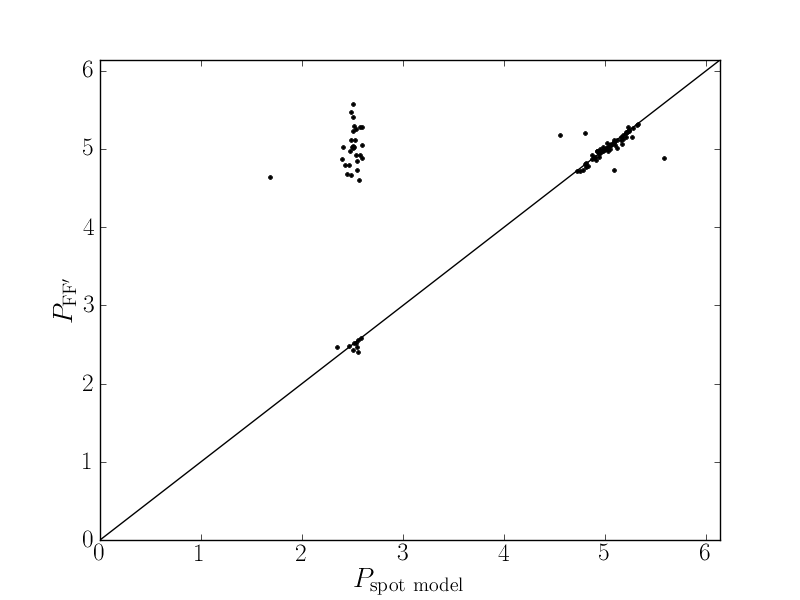}
  \includegraphics[width=0.3\linewidth]{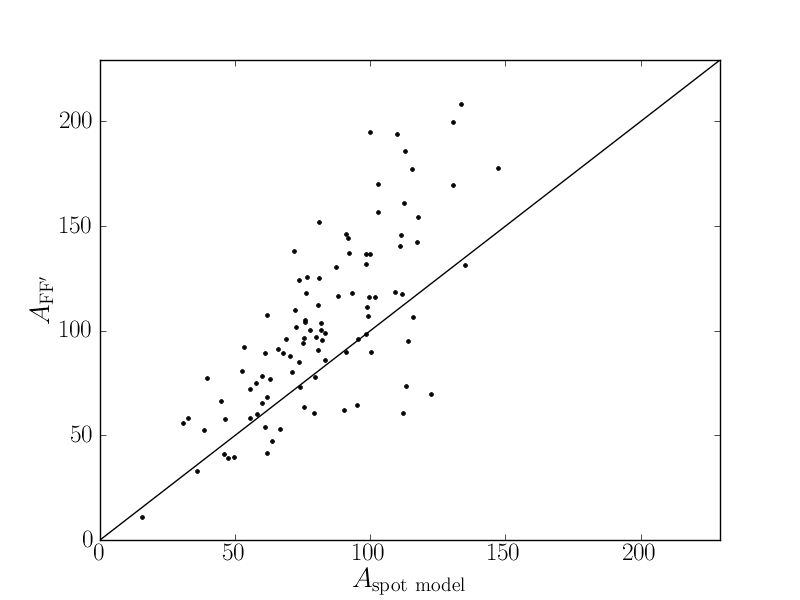}
  \caption{Comparison of RV variations simulated with our spot model (x-axis) to the predictions of the $FF'$ method applied to the photometric output of the same simulations (y-axis). The left, middle and right column show the time-series RMS scatter, the period and the amplitude of the best-fitting sinusoidal modulation, while the top and bottom rows correspond to 20 and 200 spots, respectively. The black line in each panel follows $y=x$.}
  \label{fig:simres_scatter}
\end{figure*}

We used the spot model described in Section~\ref{sec:spot_model} to
generate flux, RV and bisector span time series and test the ability
of the $FF'$ method to recover the RV signal based on the flux
information only.

We generated 100 time-series lasting 100 days each, each with 20
randomly distributed, evolving, differentially rotating spots, and
another set of 100 time-series with 200 spots. We computed photometric
and RV time-series lasting 100 days, with one point every
0.05\,day. For each realisation, the stellar inclination was drawn at
random from a distribution uniform in $\cos(i)$, and the spots were
distributed uniformly in $\sin(\delta)$, and rotated differentially
depending on their latitude, following $P(\delta) = P_0 + 0.1
\sin(\delta)$, with $P_0=5$\,days in all cases. The angular size
$\alpha$ of each spot followed a squared exponential growth and decay,
with $e$-folding times drawn from a log-normal distribution with mean
$0.4 \log P_0$ and standard deviation 0.15. The peak times were drawn
from a uniform distribution over the range $[-L/2,3L/2]$, where $L$
was the light curve duration.

The purpose of these simulations is to identify fundamental
limitations of the method, rather than to evaluate its performance in
fully realistic conditions. Therefore, we did not add noise to the
simulated data. Each photometric time-series was processed with the
$FF'$ method, as outlined in Section~\ref{sec:practical}, to generate
synthetic RVs. As no noise was added, no smoothing was necessary. The
resulting RVs are compared to the output of the spot model for a
subset of the simulations in Figure~\ref{fig:simres_comp}. There is
generally very good agreement between the time series, but the $FF'$
RVs occasionally depart from the output of the spot model by up to a
third of the peak-to-peak amplitude. In those cases, the periodograms
(which otherwise show excellent agreement) show a different ratio
between the odd- and even-numbered \textcolor{red}{harmonics}. This occurs when the
spot distributions approximate odd-numbered multipoles: as noted in
the previous section, this causes RV variations with virtually no
counterpart in photometry. In other words, as expected, RV variations
at the fundamental and first harmonic are well reproduced by the $FF'$
output and strongly suppressed in the residuals (often by as much as
80\%), but variations at higher, \textcolor{red}{even-numbered harmonics} remain, and can even be enhanced.

We also computed and compared the overall RMS of the `true' (spot-model) RVs, the $FF'$output, and the residuals, and the best fit sinusoidal period and amplitude in the $FF'$ output compared to the true RVs. The RMS of the residuals is reduced by $>50\%$ compared to the original value in 54\% of the simulations with 20 spots, but this performance is achieved in only 16\% of the simulations with 200 spots. This is because, when there are so many spots, the spot distribution almost always has an odd-numbered multipole component. Nonetheless, as shown in Figure~\ref{fig:simres_scatter}, the RMS, amplitude and period of $FF'$ output are always well-correlated with the corresponding values for the `true' RVs. For simulations using 20 spots, the Pearson correlation coefficients for the RMS and amplitudes are 0.90 and 0.84, respectively, with slightly lower values of 0.80 and 0.75 for 200 spots. There is a slight tendency for the $FF'$ to over-predict the amplitude of the variations: linear fits to the scatter plots shown in Figure~\ref{fig:simres_scatter} yield slopes of 1.08 and 1.06 (1.1 and 1.17) for the RMS and amplitude respectively, for 20 (200) spots. The period, or its first harmonic, is almost always correctly recovered (up to a precision of 10\%). Incidentally, the fraction of the cases where the first harmonic dominates over the fundamental is smaller for the $FF'$ output than for the direct spot-model simulations.

Thus, except in particularly favourable cases, the $FF'$ method does
not enable a precise `correction' of the RV variations due to
activity, prior to searching for low mass planets, for
example. However, it does permit a statistical comparison in terms of
amplitudes and frequency content. Nonetheless, further tests on real
data are desirable to establish the performance of the $FF'$ method on
a firmer footing.

\subsection{Application to HD\,189733}
\label{sec:hd189}

\begin{figure}
  \centering
  \includegraphics[width=\linewidth]{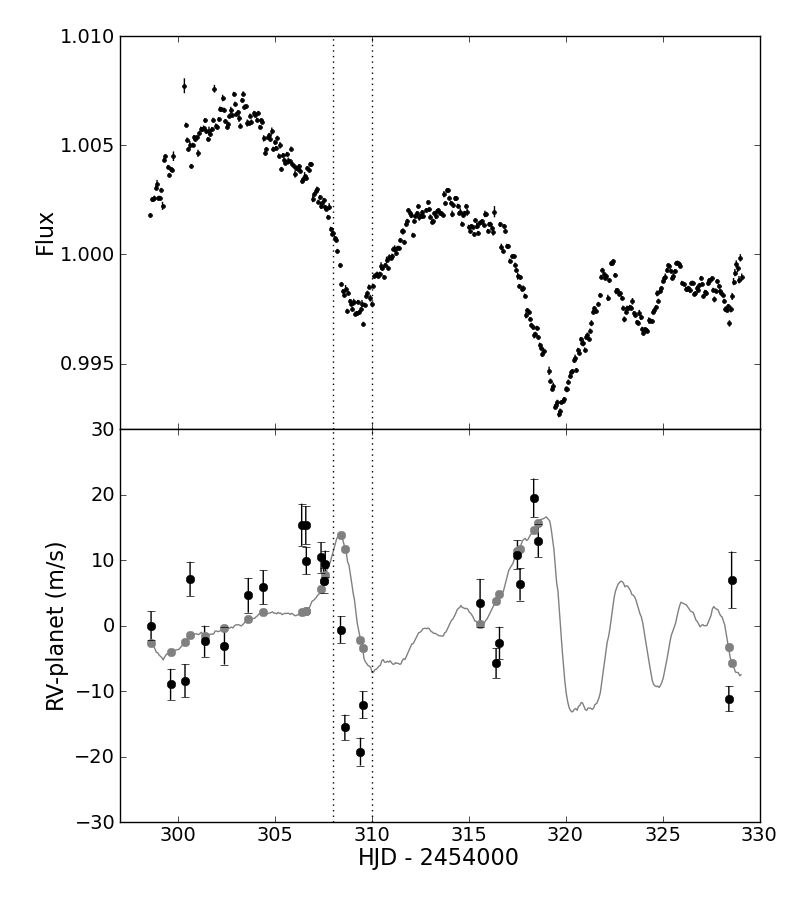}
  \caption{Photometry and RV time-series for HD\,189733. The MOST light curve \protect\citep{boi+09} is shown in the top panel and the observed and simulated RV data are compared in the bottom panel. The black dots with error bars show the SOPHIE data from \protect\citet{lan+11} after removal of the best-fit planetary signal, and subtraction of a constant 21.6\,m\,s$^{-1}$ offset. The grey line shows the RV curve simulated by applying the $FF'$ method to the MOST light curve, and the grey dots show the same curve linearly interpolated to the sampling of the SOPHIE observations.}
  \label{fig:hd189t}
\end{figure}

The transiting planet host star HD\,189733 was the target of intensive
simultaneous monitoring with the RV spectrograph SOPHIE and the
photometric satellite MOST. An in-depth analysis of these observations
from the activity point-of-view was already presented in
\citet{boi+09}. This dataset constitutes a useful test for RV jitter
simulation methods based on photometry, and was recently used for this
specific purpose by \citet{lan+11}.

Starting from the MOST light curve, which is shown in the top panel of
Figure~\ref{fig:hd189t}, we simulated the expected activity-induced RV
variations using the $FF'$ method. \textcolor{red}{The light curve was first
  smoothed using the iterative non-linear filter of \citet{aig+04} using
  a baseline of 6 data points ($\sim 10$\,h) to reduce the noise on
  the time-derivative estimate.} The results are shown as the grey
line in the bottom panel of Figure~\ref{fig:hd189t}. We then compared
this to the SOPHIE observations, which are shown as black dots. Note
 that we used the new reduction of the SOPHIE data, as described by
\citet{lan+11}, and worked with the residuals of the planetary orbit
(I.\ Boisse, priv.\ comm.). Following \citet{lan+11}, we subtracted a
constant offset of 21.6\,m\,s$^{-1}$ from the SOPHIE orbit residuals.  We then
linearly interpolated the $FF'$ output to the sampling of the SOPHIE
observations (grey dots in Figure~\ref{fig:hd189t}).

The interpolated $FF'$ output is a good match to the orbit residuals
except from ${\rm HJD}=2\,454\,308$ to $2\,454\,310$, and is virtually
identical to the \citet{lan+11} results throughout. The latter already
noted that their model could not reproduce the very rapid drop
observed in the RVs around this time. One possible explanation may be
that the spot distribution around this time had a significant
odd-numbered multipole component, which no photometry-based method
could recover. However, we note that, around ${\rm HJD}=2\,454\,308$,
the observed flux also dips faster than can be reproduced by an
unevolving surface feature rotating into view. This suggests that
there are rapidly evolving active regions on the star at this time, a
situation which neither the $FF'$ method nor the method of
\citet{lan+11} are well suited for.

\textcolor{red}{The reduced $\chi^2$ of the SOPHIE orbit residuals
  (excluding the problematic interval from ${\rm HJD}=2\,454\,308$ to
  $2\,454\,310$) is 14.45, and their r.m.s is 9.4\,m\,s$^{-1}$. Subtracting
  the activity contribution, as predicted by the $FF'$ method, reduces
  the reduced $\chi^2$ by a factor $>2$ to 6.58, and the r.m.s to
  6.6\,m\,s$^{-1}$. Although the presence of a residual activity signal cannot
  be excluded, the final r.m.s is consistent with the level of
  instrumental systematics typical of SOPHIE at the time of the
  HD\,189733 observations ($\sim 5$\,m\,s$^{-1}$, Boisse priv.\ comm.).}

\begin{figure}
 \centering
  \includegraphics[width=\linewidth]{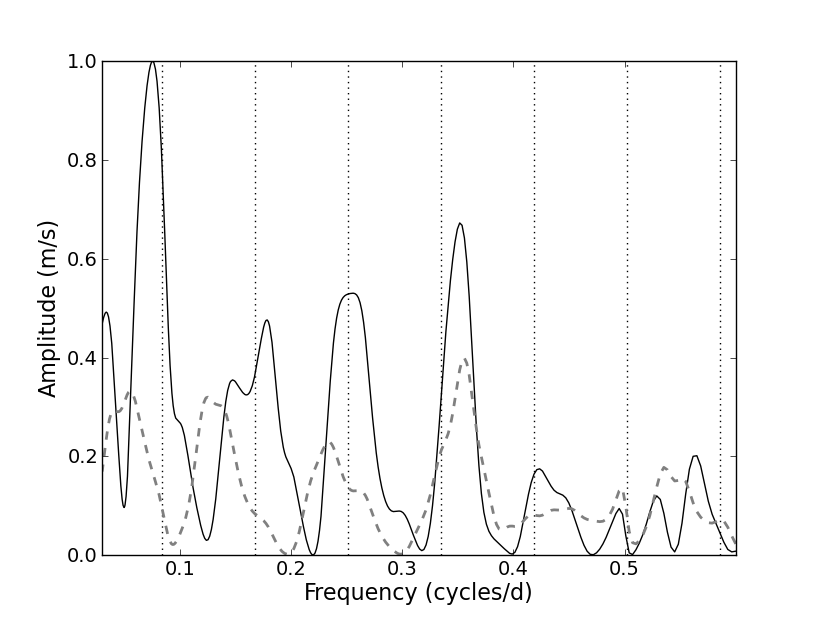}
  \caption{Lomb-Scargle periodograms of the observed RV time-series for HD\,189733 after subtracting only the planetary orbit (solid black line) and after removing the simulated activity signal also (\textcolor{red}{dashed} grey line). Both RV datasets have the same time sampling, including only SOPHIE data points simultaneous with the MOST observations, and excluding the 4 data points between the vertical dashed lines in Figure~\protect\ref{fig:hd189t}. The rotational frequency and its \textcolor{red}{harmonics} are indicated by the vertical \textcolor{red}{dotted} lines. Following \protect\citet{boi+09}, we took the rotation period to be 11.953\,days.}
  \label{fig:hd189f}
\end{figure}

We also computed the Lomb-Scargle periodogram of the RV data before
and after subtracting the simulated activity signal
(Figure~\ref{fig:hd189f}). For this calculation, we used only data
simultaneous with the MOST observations, and excluded the
aforementioned discrepant data points. Again the results are almost
identical to Figure~6 of \citet{lan+11}. Subtracting the output of the
$FF'$ method suppresses power at the rotational frequency by a factor
close to 10, and gradually decreasing fractions of the power at each
harmonic. The last significant peak (the third harmonic) is suppressed
by a factor of $\sim 2$ only.

In summary, despite its simplicity, the $FF'$ method gives results
that\textcolor{red}{, at least in this specific case,} are equivalent
to \textcolor{red}{the} more sophisticated approach \textcolor{red}{of
  \citet{lan+11}}, and achieves the same performance in terms of RV
power suppression. \textcolor{red}{One possible explanation for this
  is that the maximum entropy regularisation employed by
  \citet{lan+11}} effectively places a prior on the spatial and
temporal scales accessible to their model\textcolor{red}{. This} has a
similar effect to the \textcolor{red}{approximations} made in the
$FF'$ method\textcolor{red}{, which is only a first order
  approximation to the full expression for multiple spots, and is
  therefore expected to match the signature of the dominant active
  regions only}. Therefore, one should be cautious not to
over-interpret the output of both models, particularly as regards
short-term behaviour \textcolor{red}{ -- unfortunately, a detailed
  comparison of the two methods in the short-timescale regime is
  difficult because the RV data for HD\,189733 have relatively sparse
  time-sampling, and uncertainties comparable to the short-term
  activity signal}.

\subsection{The Sun}
\label{sec:sun}

\begin{figure*}
 \centering
  \includegraphics[width=0.48\linewidth]{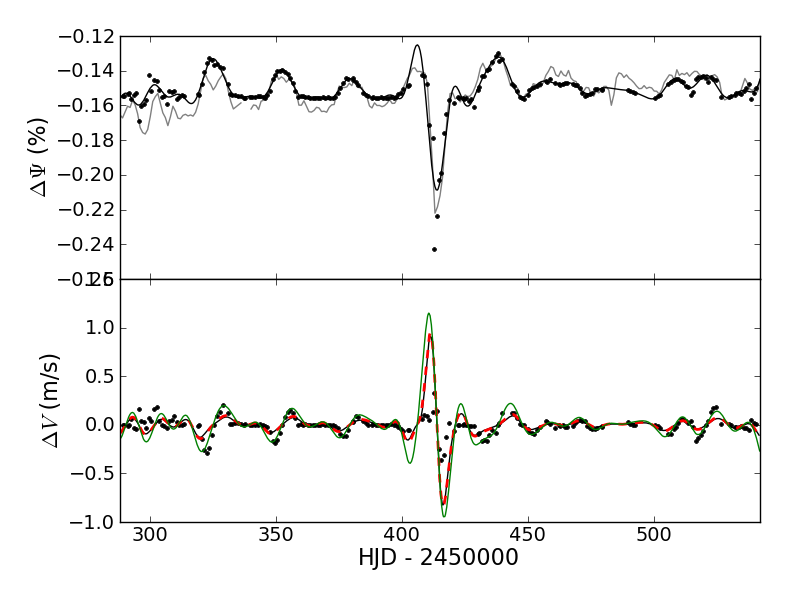}\hfill
  \includegraphics[width=0.48\linewidth]{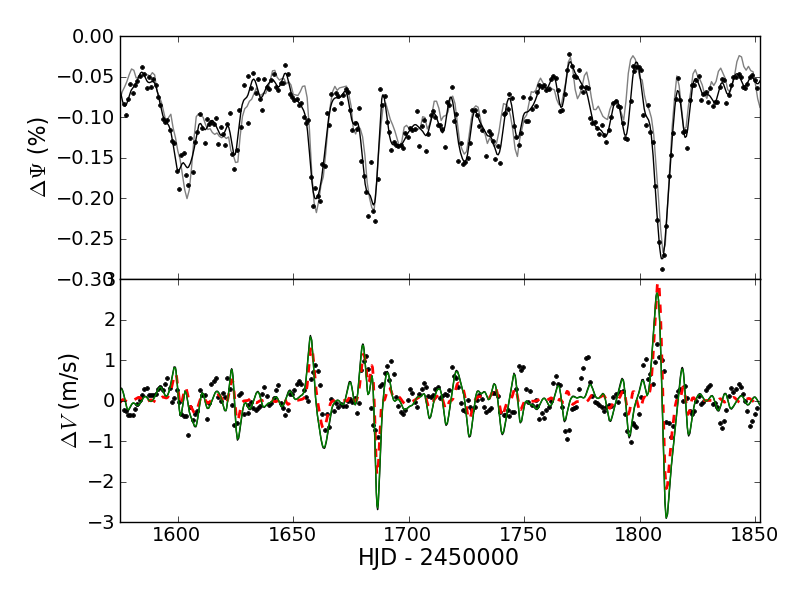} 
  \caption{Application of the $FF'$ to synthetic solar data. The black dots show the synthetic TSI (top panel) and RV (bottom panel) variations of the Sun, from \protect\citet{meu+10b}, for two 6-month periods when the Sun was relatvely inactive (left) and active (right). In each case, the solid black line in the top panel shows the smoothed version of the synthetic TSI used as input to the $FF'$. The measured TSI (SoHO/VIRGO daily average, from {\tt http://www.pmodwrc.ch/}, maintained by C.\ Fr{\"o}hlich) is also shown for comparison as the solid grey line. The solid black, thick dashed red and solid green lines in the bottom panel then show the $FF'$ predictions based on that smoothed TSI, using different values of $\Psi_0$ and $\kappa \delta V_{\rm c}$ (see text for details).}
  \label{fig:sun}
\end{figure*}

The Sun, as the nearest and best-studied star available to us, is an ideal test case for the $FF'$ method. Its brightness and RV variations have been intensively monitored for decades,  from space (e.g. by the SoHO satellite) and from the ground (e.g. by the GONG and BiSON networks). However, solar RV monitoring projects are primarily intended to study the Sun's 5-minute oscillations, and we were not able to find a publicly available set of full-disk RV measurements that were calibrated over long timescales. 

We therefore used the synthetic dataset of \citet{meu+10b}. This dataset was generated by identifying different types of magnetic features on SoHO/MDI magnetograms, and simulating their effect in both photometry and RV, tuning the model to fit the total solar irradiance (TSI) variations observed by SoHO/VIRGO. We focus on two 6-month long sections at opposite phases in the solar cycle, one during a period of relatively low activity, and one during a period of relatively high activity. For each, we were provided with synthetic TSI and RVs sampled approximately once per day (A.-M.\ Lagrange, priv.\ comm.). The synthetic RVs are, of course, subject to the limitations of the \citet{meu+10b} model. However, applying the $FF'$ to the synthetic TSI, and comparing the results to the synthetic RVs at least provides an internal consistency check. In particular, it is a fairly stringent test of the very limited treatment of faculae and plage used in the $FF'$ method, since the latter are the dominant effect in the synthetic RVs.

As can be seen in the top panel of Figure~\ref{fig:sun}, the synthetic
TSI (black dots) is a good match to the observed TSI (grey line), but
it is not smooth. Its time-sampling is also slightly irregular and
relatively sparse ($<1$ point per day). This makes smoothing using the
iterative non-linear filter discussed in Section~\ref{sec:practical},
which gave satisfactory results for the MOST time-series of HD\,189733
in Section~\ref{sec:hd189}, inappropriate for this dataset. Instead,
we performed a Gaussian process (GP) regression on the synthetic
TSI. More details on the GP regression are given in
Appendix~\ref{app:GP}.

The resulting smoothed TSI (shown as the black solid line in the top
panel of Figure~\ref{fig:sun}) was then fed in to the $FF'$ method,
and the results are compared to the synthetic RVs in the bottom
panel. We first used Equation~(\ref{eq:psi0}) to estimate $\Psi_0$,
and set $\kappa \delta V_{\rm c}$ to zero (black line). We also tried fitting
for those two parameters, using a downhill simplex optimizer to
minimize the sum of the squared differences between the synthetic RVs
and the $FF'$ output linearly interpolated to the same time-sampling
(thick red dashed line). Finally, we also tried setting $\Psi_0$ to 1,
corresponding to an absolute irradiance of $1367.62\,{\rm W}/{\rm
  m}^2$, which is approximately the maximum irradiance observed at any
time during the last solar activity cycle (green line, $\kappa V_{\rm
  c}$ was again set to zero). In the low-activity case, the red and
black lines are identical, with best-fit parameters $\Psi_0 = 0.99887$
and $\kappa \delta V_{\rm c}=12.7$\,m\,s$^{-1}$. The three versions have very similar
r.m.s. residuals, this time $\sim 0.06$\,m\,s$^{-1}$. In the high-activity
case, it is the green and black lines (and the corresponding values of
$\Psi_0$) which are virtually indistinguishable, whereas the best-fit
parameters used to produce the red curve are $\Psi_0 = 0.99926$ and
$\kappa \delta V_{\rm c}=417$\,m\,s$^{-1}$. Again, the three versions have very
similar root mean square (r.m.s.) residuals, $\sim 0.5$\,m\,s$^{-1}$. In both
the low- and high-activity cases, fixing $\Psi_0$ to 1 appears to give
a better match visually than fitting for it, but it yields a slightly
larger squared error and residual r.m.s.

The most significant deviations between the synthetic and $FF'$ RVs
occur when a large sunspot group causes a sudden drop in the TSI, e.g.\
near ${\rm HJD}-2450413$. This necessarily induces a similarly large
variation in the $FF'$ output, which has no counterpart in the
synthetic RVs of \citet{meu+10b}. There are several possible reasons
for these discrepancies: the insensitivity of the photometry to
certain spot configurations, the limited treatment of faculae and
plage in the $FF'$ method, or unidentified issues with the synthetic
RVs themselves. In the absence of \emph{measured} solar RVs to provide
a ground truth, all these possibilities must remain open.

Nonetheless, the overall agreement between the $FF'$ output and the
synthetic RVs is very encouraging. This implies that the $FF'$ method
can certainly be used to study RV variations down to the m\,s$^{-1}$ level.

\subsection{Application to Kepler data}

\begin{figure}
 \centering
  \includegraphics[width=\linewidth]{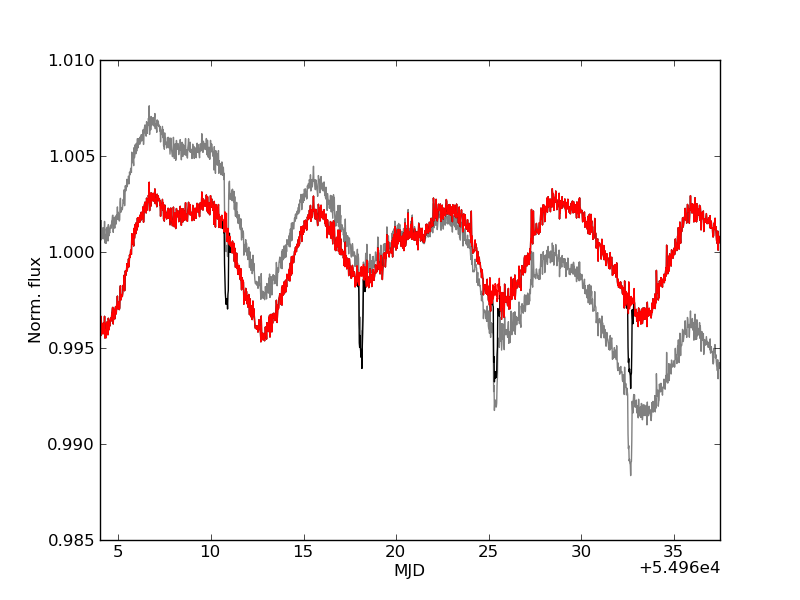} 
  \caption{Example Kepler Quarter 1 light curve before and after applying our systematics correction. The original, raw time-series is shown in grey, the corrected time-series in black, and the corrected time-series without the transits (used to estimate the RV modulations) in red. \textcolor{red}{Note that the red line completely overlaps with the black, except during the transits.} This example is KID 3642741 (KOI 242).}
  \label{fig:kepler_example}
\end{figure}

\begin{figure}
 \centering
  \includegraphics[width=\linewidth]{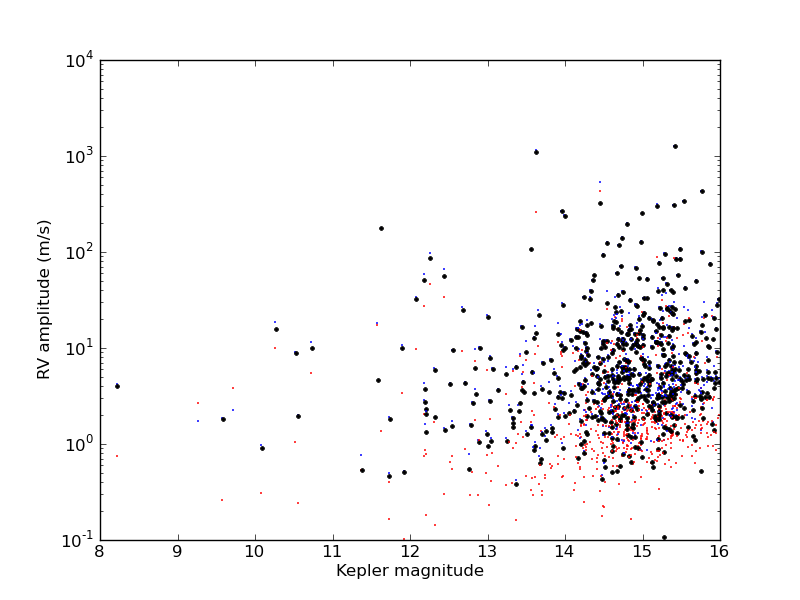} 
  \caption{Application of the $FF'$ to Kepler quarter 1 light curves containing planet candidates. The small blue dots show the RV amplitudes derived from the light curves after removing the transits and smoothing on one tenth of the dominant light curve period, while the small red dots show the RV amplitude expected for a white noise-only light curve with the same high-frequency noise level, smoothed to the same extent. The black dots show the noise-corrected RV amplitude estimates, obtaining by subtracting the latter from the former in quadrature.}
  \label{fig:kepler}
\end{figure}

Given a sample of high-precision, high time-sampling light curves, the
$FF'$ method can be used to estimate the statistics of the
activity-induced RV variability of the same sample of stars.  The tens
of thousands of high-precision light curves produced by the Kepler
mission constitute an ideal dataset to do this. A systematic
application of the $FF'$ method to individual Kepler light curves is
beyond the scope of this paper, but as an illustration we now proceed
to apply it to a subset of the light curves in which the Kepler team
identified transiting planet candidates \citep{bor+11}.

The Kepler photometric pipeline produces two versions of the light
curves: a `raw' time-series, and a version corrected for most of the
systematic instrumental effects. In the current version of the
pipeline, this correction unfortunately also removes much of the
intrinsic variability of the target stars. In the context of a
separate study, focussed on the statistics of photometric variability
in Kepler data, we have developed a more conservative systematics
removal correction, which is designed to preserve astrophysical
signals. This correction will be described in detail in a
forthcoming paper, so we only summarise the underlying principles
here. Each light curve is decomposed into a linear combination of all
the other light curves, plus an intrinsic component, using Bayesian
linear regression. The most significant trends that are common to many
light curves are identified using an information entropy criterion. They are then
combined using principal component analysis, de-composed into their
intrinsic oscillatory modes, and removed from each individual light
curve, again using Bayesian linear regression. The process is repeated
iteratively until no further trends are identified. The correction is
currently available only for the data from Quarter 1, which were
released to the public on June 15, 2010. We therefore focus on 601 of
the 997 stars with planet candidates announced by \citet{bor+11},
whose light curves were included in that release, and treat only the
Quarter 1 data, which lasts 33 days. An example of the systematics
correction applied to one of these objects is shown in
Figure~\ref{fig:kepler_example}.

After masking the transits, we first computed the periodogram of each
light curve to identify the dominant periodicity, restricting the
search to the range $0.5$--$50$\,days. The periods were not checked
individually, and there is no guarantee that they are related to a
real, physical period (such as the rotation period of the target
star). However, we use them to smooth the light curves, by applying
the iterative non-linear filter of \citet{aig+04}, with a smoothing
time-scale equal to one tenth of the identified period. We then apply
the $FF'$ method to the smoothed light curve. The results are, of
course, affected not only by the intrinsic variability of the host
star, but also by any photometric noise still present in the smoothed
light curve. The latter depends not only on the stars' magnitude, but
also on the smoothing timescale used, which varies from star to
star. This is unavoidable -- excessive smoothing of a light curve with
real, short-term variability would lead to under-estimated RV
variations -- but it is important to quantify the residual noise
contribution to the RVs. In each case, we therefore also computed the
RV variations expected for a simulated light curve containing only
white Gaussian noise at the same level as the original, smoothed using
the same timescale. The noise level was estimated as the standard
deviation of the original light curve minus the smoothed version
thereof. The noise-only RV amplitude was subtracted from the amplitude
derived from the real light curve, yielding an estimate corrected for
photometric noise.

Figure~\ref{fig:kepler} shows the resulting RV peak-to-peak amplitudes
over the duration of Quarter 1, as a function of Kepler magnitude. The
small blue dots show the amplitude measured for the real light curves,
and the small red dots the corresponding amplitudes for noise-only
light curves, and the black dots show the noise-corrected
estimates. The distribution of the red points shows that the Kepler
light curves can be used to estimate RV variations down to the 2--3\,m\,s$^{-1}$
level, below which they become noise-dominated in most cases. The
distribution of the noise-corrected amplitudes is approximately
log-normal, with a mean of $\sim 4.1$\,m\,s$^{-1}$ and a width of $\sim
0.45$\,dex. Approximately two thirds of the candidates are expected to
display RV variations significantly above the noise limit. This does
not translate directly into implications for the detectability of the
RV signatures of the transit candidates, as the intrinsic RV
variations may occur on timescales which are quite distinct from the
orbital period. Nonetheless, it does suggest that RV confirmation
would be challenging for these candidates, the majority of which have
radii below that of Neptune (for comparison, the expected
RV-semi-amplitude for a Neptune-mass planet in a 10 day orbit around a
Sun-like star is $\sim 5$\,m\,s$^{-1}$). The predicted amplitude of the
activity-induced RV signal is below the noise level for the remaning
third of the candidates. 

\section{Conclusions}
\label{sec:conc}

We have presented a new method to derive the stellar variability
expected in RV measurements from well-sampled light curves, which
requires no knowledge of the rotation period or detailed spot
modelling. We call this method $FF'$ because it uses the light curve
and its first derivative. \textcolor{red}{A number of approximations are
  built into our method: in particular, it ignores limb-darkening, as
  well as the photometric effect of faculae (but not their RV
  signature), and the expression used to compute the RV signal is
  accurate only to first order when multiple spots are present on the
  stellar surface. However, the observables (photometry and RV) are disk-integrated
  quantities, which mitigates the impact of these approximations.}

We tested this $FF'$ method on time-series simulated using the simple
spot model on which it is loosely based. Overall, our simulations show
that the RMS, period and amplitude of periodic modulation of $FF'$
predictions are in fairly good agreement with the corresponding model
values. Given its simplicity, our method reproduces activity-induced
variability surprisingly well, matching the latter's amplitude and
distribution of power at the \textcolor{red}{first few harmonics of the rotational frequency} to $\sim 10\%$ or better in most cases. There are
exceptions, where the power at \textcolor{red}{odd-numbered harmonics} is not
well-matched. To understand this phenomenon, we used our spot-model to
explore the relationship between photometric and RV variations due to
spots in the general sense, and showed that any method which uses
photometry to simulate RVs will struggle to reproduce the signals if
spot distribution has a significant odd-numbered multipole component.

We then applied the $FF'$ method to three example datasets: the
benchmark SOPHIE/MOST observations of HD\,189733, synthetic TSI and RV
data for the Sun, and the Quarter 1 light curves of Kepler planet
candidate host stars.

For HD\,189733, our results are essentially identical to those
obtained by \citet{lan+11} with a much more sophisticated method,
reducing the power of the RV variations at the rotational frequency by
a factor $\sim 10$, and by a factor of a few at the first few
\textcolor{red}{harmonics}. Where both approaches struggle to
reproduce some of the observed RV variations, we identify a possible
explanation in their common reliance on the photometry to predict the
RVs.

The solar tests demonstrate the performance of the method down to the
sub-m\,s$^{-1}$ level. We show that, given enough high-quality RV data, it is
possible to fit for the un-spotted flux level, but this has only a
relatively minor impact, and the results obtained by estimating it
from the light curve itself are almost as good. We also show that it
is possible to fit for the parameter $\kappa \delta V_{\rm c}$, which
controls the importance of the convective term, but again this has
relatively little effect on the results. These tests also illustrate
the use of Gaussian processes, rather than simpler filters, to
interpolate and smooth the photometric time-series. This opens up the
possibility of applying the $FF'$ method to cases where there is only
sparse photometry, or where a photometric proxy is derived from the
same spectra as the RV observations themselves.

Finally, we have shown that the $FF'$ method can be used to estimate
RV amplitudes down to the 2--3\,m\,s$^{-1}$ level using Kepler
data. Approximately two thirds of the 600 planet candidate host
stars to which we applied it are expected to display peak-to-peak RV
amplitudes over 30-day timescales that are above that level. This
confirms that obtaining mass measurements for many of the Kepler
terrestrial and/or longer-period planet candidates will be
challenging. However, it also opens up a range of possibilities for
identifying the most promising candidates to follow up, and perhaps
disentangling activity- and planet-induced RV signals for those
objects. We caution that, like all methods based on photometry, some
care must be taken when interpreting the output of the $FF'$ method
for individual objects. Nonetheless, we have shown that it performs
as well in that respect as other methods which have already been put
to that use.

The distinguishing characteristic of the $FF'$ method, however, is its
simplicity, speed, and lack of free parameters. In its simplest
implementation, all it requires is an estimate of the stellar radius.
The method can thus be applied to large samples of light curves such
as those produced by the CoRoT and Kepler missions, to assess the RV
jitter properties of a large sample of stars, and their implications
for the detection of habitable planets by RV. No other method we are
aware of can be applied to a large enough sample of light curves to
undertake such a project, which would form a natural extension to the
present paper.

\section*{Acknowledgments}

We are grateful to I.\ Boisse for kindly providing the MOST and SOPHIE data of HD\,189733b, to J.\ Rowe and J.\ Matthews for allowing us to use the MOST light curve, and to A.-M.\ Lagrange for providing the synthetic solar data. Some of the code used in this work was written by N.\ Gibson. This work was supported in part by the UK Science and Technology Facilities Council via standard grant ST/G002266/1 (SA) and an advanced fellowship (FP) , and by the Israel Science Foundation / Adler Foundation for Space Research via grant 119/07 (SZ). The authors wish to thank H. Kjeldsen, A.-M. Lagrange and the participants of the 2009 Exeter workshop \emph{Unsinkable planets: telluric planet detection in the presence of stellar activity} for useful discussions. Finally, we would like to thank the referee, A.\ Lanza, for his careful reading of the manuscript and helpful suggestions for improvement.

\bibliographystyle{mn2e} \bibliography{ffprime}
\bsp

\appendix

\section{Calculations underlying the spot model}
\label{sec:append}

\subsection{Trajectory of the spot}

Consider a spot located at latitude $\delta$ on the surface of a star with radius $R_\star$ rotating with period $P_{\rm rot}$ and inclination $i$ (where $i$ is defined as the angle between the star's rotation axis and the line-of-sight). We define two reference frames $S$ and $S'$, sharing the same origin, which coincides with the centre of the star, and the same $y$-axis. The stellar rotation vector defines the $+z$-direction of frame $S$, whilst the $x$-axis of frame $S'$ points towards the observer. In frame $S$, the location of the spot is given by:
\begin{displaymath}
  \frac{1}{R_\star}
  \left[
    \begin{array}{l}
      x \\
      y \\
      z 
    \end{array} 
  \right]
  = 
  \left[
    \begin{array}{l}
      \cos \delta \cos \phi \\
      \cos \delta \sin \phi  \\
      \sin \delta
    \end{array} 
  \right],
\end{displaymath}
where $ \phi = 2 \pi t / P_{\rm rot} + \phi_0 $ and $\phi_0$ is the longitude of the spot at $t=0$. The coordinates of the spot in frame $S'$ are then obtained by performing a rotation by angle $-(\pi/2-i)$ about the $y$ axis:
\begin{displaymath}
  \begin{array}{rcl}
    \frac{1}{R_\star}
    \left[
      \begin{array}{l}
        x' \\
        y' \\
        z' 
      \end{array} 
    \right]
    & = & 
    \left[
      \begin{array}{lll}
        \sin i & 0 & \textcolor{red}{+}\cos i \\
        0 & 1 & 0 \\
        \textcolor{red}{-} \cos i & 0 & \sin i \\
      \end{array} 
    \right] 
    \left[
      \begin{array}{l}
        \cos \delta \cos \phi(t) \\
        \cos \delta \sin \phi(t)  \\
        \sin \delta
      \end{array} 
    \right]
    \\
    & = & 
    \left[
      \begin{array}{l}
        \cos \delta \cos \phi(t) \sin i\ \textcolor{red}{+} \sin \delta \cos i \\
        \cos \delta \sin \phi(t)  \\
        \textcolor{red}{-} \cos \delta \cos \phi(t) \cos i + \sin \delta \sin i
      \end{array} 
    \right].
  \end{array}
\end{displaymath}

\subsection{Relative drop in flux due to the spot}

To calculate the quantities of interest, we need to evaluate $\beta$, the angle between the spot normal and the line of sight. It is easy to show (for example using the cosine rule on the triangle defined by the spot's position vector and its projection onto the line-of-sight) that this angle is simply given by
\begin{displaymath}
  \cos \beta = x' / R_\star = \cos \delta \cos \phi(t) \sin i + \sin \delta \cos i.
\end{displaymath}
The relative drop in observed flux $F$, described by equation~(\ref{eq:F}), follows the same time dependence.

In frame $S$, the vector describing the rotational motion of the stellar surface at the location of the spot is given by
\begin{displaymath}
  V_{\rm rot} = 
  \left[
    \begin{array}{l}
      \textcolor{red}{-} V_{\rm eq} \cos \delta \sin \phi \\
      V_{\rm eq} \cos \delta \cos \phi \\
      0
    \end{array}
  \right]
\end{displaymath}
where $V_{\rm eq} = 2 \pi R_\star / P_{\rm rot}$. In frame $S'$ this becomes
\begin{displaymath}
  \begin{array}{rcl}
    V'_{\rm rot}
    & = & 
    \left[
      \begin{array}{lll}
        \sin i & 0 & - \cos i \\
        0 & 1 & 0 \\
        \cos i & 0 & \sin i \\
      \end{array} 
    \right] 
    \left[
      \begin{array}{l}
        \textcolor{red}{-} V_{\rm eq} \cos \delta \sin \phi \\
        V_{\rm eq} \cos \delta \cos \phi \\
        0
      \end{array}
    \right]
    \\
    & = & 
    \left[
      \begin{array}{l}
        \textcolor{red}{-} V_{\rm eq} \cos \delta \sin \phi \sin i \\
        V_{\rm eq} \cos \delta \cos \phi \\
        V_{\rm eq} \cos \delta \sin \phi \cos i \\
      \end{array} 
    \right].
  \end{array}
\end{displaymath}
The time dependence of the RV signature of the spot, $\delta RV_{\rm s}$, is governed by the $x$-component of this vector times $F(t)$.

The net convective up-welling velocity is radial, so its line-of-sight component is simply:
\begin{displaymath}
  \delta V'_{\rm c} =  \delta V_{\rm c} \cos \beta.
\end{displaymath}

\section{Gaussian process regression on the total solar irradiance}
\label{app:GP}

In order to apply the $FF'$ method to the solar data, we need to convert the irregularly sampled, discontinuous synthetic TSI to a smooth, tightly sampled flux estimate. We do this using Gaussian process (GP) regression. 

GP models are extensively used in the machine learning community for Bayesian inference in non-parametric regression and classification problems. By definition, any vector of observations drawn from a Gaussian process has a joint distribution which is a multi-variate Gaussian. Families of random functions which share the same smoothness and covariance properties can be parametrised in terms of  a covariance function or kernel, which specifies the covariance between pairs of points as a function of -- typically -- the distance between these points in some input space (e.g. the time interval between two observations). Thus, GP regression is particularly well-suited to modeling time-series containing correlated stochastic signals and noise.  A brief introduction to GP regression, with an application to space-based, high-precision stellar photometry, albeit in a different context (exoplanet transmission spectroscopy), can be found in \citet{gib+11}. The textbook by \citet{ras+06}\footnote{Available online at {\tt www.GaussianProcess.org/gpml}.} provides a more detailed treatment of GPs for both regression and classification. 

The most important step in GP regression is the choice of covariance function, or kernel. Most kernels are decreasing functions of the distance between pairs of points in the input space. One of the simplest and most widely used kernels is the squared exponential:
\begin{equation}
k_{\rm SE}(t,t') = \theta^2 \exp \left( - \frac{r}{2 \tau^2} \right).
\end{equation}
where $k_{\rm SE}(t,t')$ is the covariance between observations taken at times $t$ and $t'$, and $r \equiv |t-t'|$ is the time interval between the two observations, $\theta$ is a parameter controling the amplitude of the flux variations, and $\tau$ is a parameter controlling the time-scale of the flux variations. 

Visual examination of the synthetic solar TSI indicated that the data might contain variations on more than one time-scale. Such behaviour can be modelled using a rational quadratic kernel:
\begin{equation}
k_{\rm RQ}(t,t') = \theta^2 \left( 1 + \frac{r}{\alpha l^2} \right)^{-\alpha} 
\end{equation}
where $\alpha$ is an additional parameter, controling the distribution of timescales. Indeed, it can be shown \citep{ras+06} that the rational quadratic kernel is equivalent to a superposition of an infinite number of squared exponential kernels with a distribution of time-scales $\tau$ that is a power-law of index $-\alpha$. When $\alpha \rightarrow \infty$, the rational quadratic kernel approximates the squared exponential kernel. For finite $\alpha$, the rational quadratic implies significantly more covariance at relatively large separation than the squared exponential (equivalent to a `long tail' behaviour). 

Stellar light curves, which display the effects of rotationally modulated, evolving active regions, tend to be quasi-periodic. This kind of behaviour can also be modelled by a GP, using a periodic kernel multiplied by a squared exponential term:
\begin{equation}
k_{\rm QP} = \theta^2 \exp \left( - \frac{\sin^2(\pi r / P)}{2 T^2} - \frac{r}{2 \tau^2} \right).
\end{equation}
where $P$ is the period in days, $T$ is the time-scale of variations within a period, and $\tau$ is now the evolution time-scale (also in days), controling the rate of change of the shape and amplitude of the signal from one period to the next. We could also have combined the periodic term with a rational quadratic term, but initial experiments with that possibility indicated that this gave too much freedom to the model.

These are only some of the possible kernels which are relevant to the
type of dataset we are trying to model here, see \citet{ras+06} for a
more detailed discussion of covariance functions. For a given set of
kernel parameters, the GP defines a probability distribution over
functions sharing the same covariance properties. This distribution
can be conditioned on any available observations, yielding a
predictive distribution which can be used to interpolate the data to
the desired sampling. The process also yields a marginal likelihood,
which can be maximised with respect to the kernel parameters (which
are also known as the hyper-parameters of the GP) in order to optimise the latter.

We experimented with all the kernels listed above on the synthetic
solar TSI, adding a delta-function to represent white noise:
\begin{equation}
k_{\rm tot}(t,t') = k(t,t') + \theta_{\rm w} \delta(r).
\end{equation}
In fact, the synthetic data we are \textcolor{red}{modeling} contain no explicit white
noise term, but the addition of a small constant noise term to the
diagonal elements of the covariance matrix significantly helps
convergence.

In the high-activity case, the squared exponential kernel yielded a
poor fit to the data (low marginal likelihood after optimising the
hyper-parameters): a single time-scale is not sufficient to model the
data. The quasi-periodic kernel also gave a relatively low marginal
likelihood, with a short period (unrelated to the known solar rotation
period), $T \gg 1$ and $\tau \ll P$ \textcolor{red}{(note that, unlike $\tau$, $T$ is dimensionless, because it divides the $\sin^2$ term, which varies between 0 and 1)}. This implies that the active
regions are evolving on timescales comparable to, or shorter than, the
rotation period. With such a combination of hyper-parameters, the
quasi periodic GP behaves essentially like the squared
exponential. The best results were obtained with the rational
quadratic kernel, with $\theta = 0.0004$, $\alpha = 1.5$, $\tau =
2.9$\,days and $\theta_{\rm w} = 0.0001$ respectively \textcolor{red}{($\theta$ and $\theta_{\rm w}$ are in units of relative flux and $\alpha$ is dimensionless)}. The mean of the
predictive distribution obtained with these hyper-parameters was used
as the smoothed TSI.

In the low-activity case, the best results were obtained with the
squared exponential kernel, with $\theta = 0.001$, $\tau = 12.7$\,days
and $\theta_{\rm w}=0.00005$. When using the rational quadratic
kernel, the best-fit value of $\alpha$ was very large -- which
approximates the behaviour of a squared exponential kernel. When
excluding the sunspot crossing around ${\rm HJD} = 2450413$, the quasi
periodic kernel gave relatively good results with best-fit hyper
parameters $\theta = 0.0008$, $P = 28.9$, $T=1.4$, $\tau = 48.3$ and
$\theta_{\rm w} = 0.00002$. However, this kernel could not reproduce
the TSI during the aforementioned sunspot crossing, so we used the
squared exponential to generated the smoothed TSI.

\label{lastpage}

\end{document}